%% file: main.tex
\documentclass[conference]{IEEEtran}
\IEEEoverridecommandlockouts

\usepackage{amsmath,amssymb,amsfonts}
\usepackage{graphicx}
\usepackage{textcomp}
\usepackage{enumitem}
\usepackage{balance} 
\usepackage{xcolor}
\usepackage[numbers]{natbib}
\usepackage[bookmarks=true]{hyperref}

\usepackage{comment}
\usepackage{hhline}
\usepackage{caption}
\usepackage{subcaption}
\usepackage{amsmath}
\usepackage{tikz}
\usetikzlibrary{positioning}
\usepackage[export]{adjustbox}
\usepackage{pgfplots}
\usepackage{pgfplotstable}
\pgfplotsset{compat=1.7}
\usepgfplotslibrary{groupplots}
\usepackage{multirow}

\makeatother

\usepackage{fancyhdr}

\begin{document}

\title{\LARGE \bf
What Predicts Interpersonal Affect? \\
Preliminary Analyses from Retrospective Evaluations
}


\author{\IEEEauthorblockN{
    \parbox{\linewidth}{\centering
    Maria Teresa Parreira, Michael J. Sack, Malte Jung}
    }
    \IEEEauthorblockA{\textit{Cornell University}}}

\maketitle
\thispagestyle{fancy}

\begin{abstract}

While the field of affective computing has contributed to greatly improving the seamlessness of human-robot interactions, the focus has primarily been on the emotional processing of the self, rather than the perception of the other. To address this gap, in a user study with 30 participant dyads, we collected the users' retrospective ratings of the interpersonal perception of the other interactant, after a short interaction. We made use of CORAE, a novel web-based open-source tool for \textit{COntinuous Retrospective Affect Evaluation}. In this work, we analyze how these interpersonal ratings correlate with different aspects of the interaction, namely personality traits, participation balance, and sentiment analysis. Notably, we discovered that conversational imbalance has a significant effect on the retrospective ratings, among other findings. By employing these analyses and methodologies, we lay the groundwork for enhanced human-robot interactions, wherein affect is understood as a highly dynamic and a context-dependent outcome of interaction history.
\end{abstract}


\IEEEpeerreviewmaketitle

\input{sections/00_introduction}

\input{sections/02_CORAE}
\input{sections/03_experimentaldesign}
\input{sections/04_results}

\input{sections/05_discussion}



\section*{Acknowledgment}
We thank Nawid Jamali and Hifza Javed for the collaboration on CORAE and user study design. This work was supported by Honda Research Institute USA, Inc.. 

\small
\bibliographystyle{abbrvnat}
\balance
\bibliography{bibliography.bib}

\end{document}

%% file: sections/00_introduction.tex
\section{Background}

Research in human-robot interaction is often focused on measuring users' external outcomes (e.g., improve group task performance \cite{Tennent2019}) or internal individual states (e.g., sensing student engagement \cite{bourguet2020engagement}). However, for robots to both understand and shape interactions among humans, they require an understanding of internal interpersonal states -- that is, the perception of the other through the user's perspective \cite{jung2017affective}.

Observable behavior and subjective experience are highly dynamic \cite{kuppens2017emotion}. In interactions, these aspects co-evolve over time for all the interactants \cite{butler2013emotional}, interacting in ways that are yet to be fully explored. To study these dynamics, continuous representations of affective states have been popularized \cite{gunes2013trends,mettallinou2013review}, allowing for an understanding of how humans aggregate affect information across time and unveiling regions of ``emotional saliency'', which may be pivotal to assessing the emotional experience \cite{mettallinou2013review}. 

A popular approach for collecting continuous affect data is retrospective analysis \cite{cowie2007feeltrace,cowie2013gtrace,melhart2019pagan,lopes2017ranktrace}, which relies on the phenomenon that individuals often re-experience emotions when reliving a situation \cite{gottman1985valid}. In prior work \cite{2023CORAE}, we introduced CORAE, an intuitive tool for \textit{\textbf{CO}ntinuous \textbf{R}etrospective \textbf{A}ffect \textbf{E}valuation}. This tool enables researchers to collect continuous affect data about interpersonal perceptions. Participants retrospectively rank how another interactant came across immediately following an interaction, thus allowing us to capture interpersonal affective perceptions rather than feelings or affective state inferences. In other words, our system allows us to capture data about how people perceive each other emotionally continuously over time.

While affect data has been collected across subjective experience \cite{ruef2007continuous, csikszentmihalyi2014validity, gottman2000timing}, and observable behavior \cite{jung2016thin, coan2007specific}, we still lack continuous data about how \textit{affective perceptions of others} develop dynamically over time. Prior work identified relational emotion metrics that take temporal patterns of emotional expressions in interactions into account and that focus on relationally relevant dimensions of emotion expression. \cite{jung2012team,jung2016thin}. However, these metrics often rely on manual coding of emotional expressions, while automated approaches for relationally relevant emotion metrics remain to be developed.

 \begin{figure}[t] 
    \centering
    \includegraphics[width=0.43\textwidth]{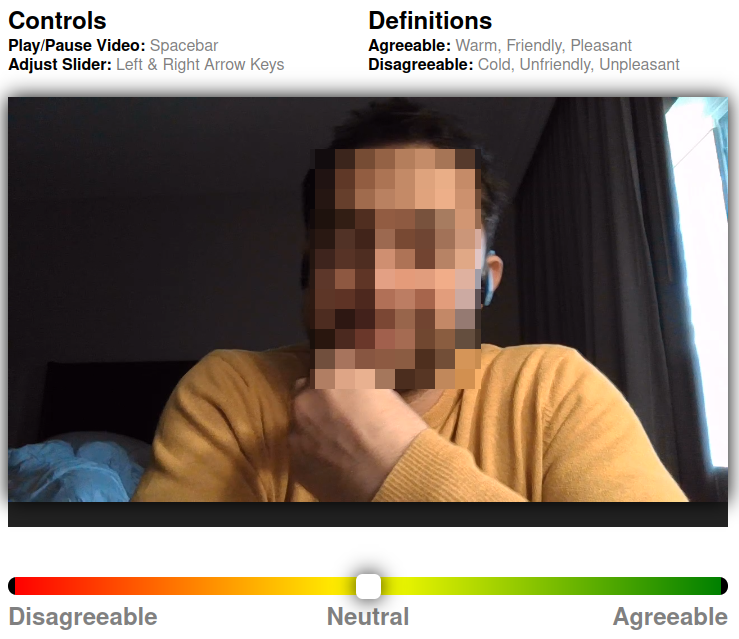}
    \caption{Annotation dashboard for CORAE. Participants retrospectively evaluate how the other person came across by reviewing the recording of the interaction with only the video from the other participant and audio of both.}
    \label{fig:frontend}
\end{figure}

We ran an online user study where 30 participant dyads interacted while completing a task. Following their discussion, participants retrospectively annotated the interaction, evaluating \textit{how the other participant came across}. In the present work, we explored how different interaction variables -- personality of the interactants, conversational balance, emotional content of participants' speech -- affected how users perceived each other. 

Our perceptions of robots are yet to be fully understood \cite{stockhomburg2021survey}, but authors have recognized the value of robotic systems that account for the dynamic nature of affect \cite{churamani2020continual}. This preliminary analysis sheds light on aspects of interactions that impact our interpersonal states, with the potential to better inform the development of these systems.

%% file: sections/02_CORAE.tex
\section{CORAE}

We provide below a brief description of CORAE, to facilitate the understanding of the interpersonal perception data used in this study.
The CORAE platform (related to the Latin word for ``heart'') \cite{2023CORAE} enables individuals to intuitively evaluate how a person's behavior is interpreted emotionally during interactions. It is an open-source tool that can be found in \href{https://corae.org}{corae.org}.

\begin{description}[align=left, leftmargin=0em, labelsep=0.2em, font=\textbf, itemsep=0em,parsep=0.3em]

\item[Design: ]
CORAE is intuitive and visually minimal (see \autoref{fig:frontend}). The central focus is a video of the other interactant. Brief instructions above the video player describe the controls of the annotation dashboard (\textit{Spacebar} to toggle playback and \textit{Left and Right Arrows} to control the slider), as well as a brief description of the terms used for measuring interpersonal perception, which can be personalized to the platform's use case.

A progress bar is displayed below the video player to inform participants what proportion remains of their evaluation. Finally, below the video player is displayed the annotation slider. The annotation bar is bounded and discretized (a total of 15 points, from $-7$ --- \textbf{Disagreeable} --- to $+7$ --- \textbf{Agreeable}). Participants may only change their rating during video playback and are constrained by the platform to do it ``continuously" (i.e., they cannot instantaneously change the rating from \textit{Neutral} ($0$) to \textit{Agreeable} ($7$), but rather adjust to each value in sequence).

\item[Data Logging: ]
Data is logged for a session in two ways: (1) by default, the mode for data logging is set to predetermined intervals of one second; and (2) whenever a change in the rating occurs. Associated data points are the slider position (rating), time code, and video frame (in the format \textit{``SliderNumericalPosition": ``Hours:Minutes:Seconds:VideoFrame"}), which are logged in a JSON file.

\end{description}

%% file: sections/03_experimentaldesign.tex
\section{User Study}
\label{sec:study}

We carried out a user study to collect dyadic interaction data and retrospective interpersonal ratings through CORAE. A more detailed description can be found in \citet{2023CORAE}.

\subsection{Experimental Procedure}

Participants were recruited through Prolific\footnote{\url{https://www.prolific.co/}}. Before scheduling their slot, each participant read and signed a consent form. The study took place fully online, with interactions facilitated through Zencastr\footnote{\url{https://zencastr.com/}}, a video call platform that allows for the recording of each video and audio stream separately. Participants read task instructions, including a description of the discussion topic (\textit{Reasons for Poverty task} \cite{shek2002reasons}, detailed below). After this, participants were recorded while interacting to solve the task. When they reached an agreement, or after 10 minutes of discussion, participants were asked to stop discussing and fill out a survey. This survey collected demographic data, as well as measures of interpersonal affect. Each participant was then distributed a URL that opened an instance of CORAE's annotation platform in their browser. Participants were each presented with a video of their discussion partner, and the audio of both, and were asked to continuously rate how their partner came across, moment-by-moment. Once finished, participants completed an exit survey with open comments. Participants were compensated for their participation with US\$14, through Prolific. 

\begin{description}[align=left, leftmargin=0em, labelsep=0.2em, font=\textbf, itemsep=0em,parsep=0.3em]

\item[Reasons for Poverty task: ]
We used a modified version of the \textit{Reasons for Poverty} task \cite{shek2002reasons} as a discussion prompt. The task requires participants to agree on selecting 5 items from a list of ``reasons for poverty'', and rank them according to their ``accuracy''. Some options in the 10-item list include \textit{``Poor people lack the ability to manage money.''}, or \textit{``The society lacks justice''}. We aimed to elicit an emotionally engaging interaction.

Participants were given a maximum of 10 minutes to discuss, to prevent individuals from getting disengaged when reviewing their discussion on CORAE.
\end{description}

\subsection{Research Questions}
In this preliminary analysis, we wanted to investigate if aspects of the users' identity, or conversational aspects such as the balance of the participation of the users were impactful for how participants perceived each other. We focused on the following research questions:
\begin{itemize}
    \item \textbf{RQ1:} Do personality traits or demographic aspects (age, gender) of a user impact how they come across to the other interactant?
    \item \textbf{RQ2:} Do dynamics of the user's conversation (imbalance, duration of conversational turns) impact the users' interpersonal perceptions of each other?
    \item \textbf{RQ3:} Does the emotional content of the user's speech impact how they come across to the other interactant?
\end{itemize}

\subsection{Measures}

To answer the research questions laid out above, we defined a set of measures from the data collected during the interaction sessions. 

\begin{description}[align=left, leftmargin=0em, labelsep=0.2em, font=\textbf, itemsep=0em,parsep=0.3em]

\item[Interpersonal measures:]
We evaluated the perception of the other through two measures. The \textit{Interpersonal Agreeableness} measure was operationalized by asking participants  \textit{``How did the other participant come across?"} on a 7-point Likert scale (from \textit{disagreeable} to \textit{agreeable}) in the post-interaction survey. Additionally, \textit{Interpersonal Perception (IP)} was extracted from the continuous interpersonal rating data collected via CORAE. To calculate the IP, and in line with prior work\cite{jung2016thin, gottman1992marital}, for each participant, we took the cumulative sum of the ratings during the interaction and fitted a linear regression to that data. The Interpersonal Perception measure is given by the slope of that regression, providing an understanding of how the perception of the other interactant evolved over the interaction (see Figure \ref{fig:cir}).

 \begin{figure}[ht] 
    \centering
    \includegraphics[width=0.45\textwidth]{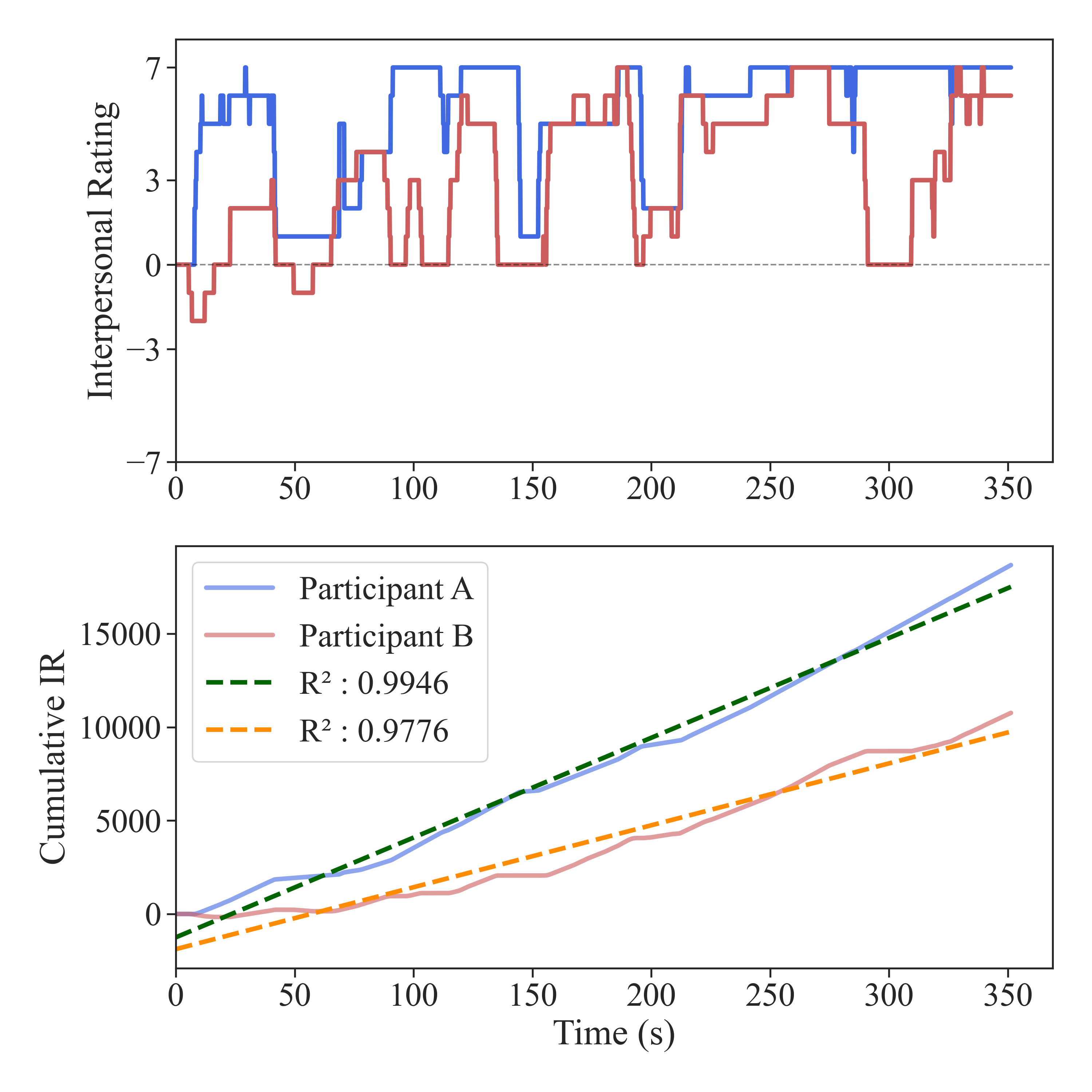}
    \caption{Dynamics of interpersonal ratings (IR) in one session sessions (top) and respective cumulative sum of rated values (bottom). Interpersonal Perception is calculated by adjusting a linear regression to the latter curves, and extracting the slope of these lines (bottom, overlaid dashed curves).}
    \label{fig:cir}
\end{figure}

\item[Demographic and Personality measures:] 
In the post-interaction survey, we collected demographic information (\textit{age, gender, nationality, race/ethnicity}) and \textit{personality traits} through the short-version of the Big Five Inventory \cite{rammstedt2007bigfive}. Participants were also asked to rate their \textit{religiousness} (\textit{not-at-all religious} to \textit{very religious}) and \textit{political leaning} (\textit{very liberal} to \textit{very conservative}) with 7-point Likert scales. 

\item[Conversational Balance measures:]
To assess \textbf{\textit{RQ2}}, we measured the users' \textit{total number} and average For \textit{duration of conversational turns}, as well as participation \textit{imbalance} \cite{Tennent2019,gillet2021balance}, which is given by:

\begin{equation}\label{eq:reward}
     \text{Imbalance} = \sum_{i\in[1,2]} |\text{s}^i - \overline{\text{s}}|
\end{equation}

where $\text{s}^i$ is the amount of time that participant $i$ has spoken over the total amount of speech. $\overline{\text{s}}$ is the mean of the relative speech time of the two participants, which in the case of a dyadic interaction is $0.5$. This measures how much the participation of the users deviated from a perfectly balanced interaction. These measures were collected from Voice Activation Detection (VAD).

\item[Speech emotion content measures:]
We transcribed the speech of the participants using the speech-to-text model Whisper \cite{radford2022robust}. Also through Whisper, we analyzed the emotional ``tone'' of each utterance of the participants in the dataset (either \textit{neutral, positive} or \textit{negative}). Because the task contained prompts that were negative, we extracted only the \textit{positive emotional tone ratio}, i.e. the ratio of utterances that were evaluated as positive (over total number of utterances). Additionally, we collected the total number of \textit{agreement words} within the speech of each user, to obtain a measure of affirmed agreement from each participant. 

\end{description}

\subsection{Participants}

To potentially elicit disagreement during the interactions, participants were selected according to their political leaning (one conservative- and one liberal-leaning). Other recruitment criteria were proficiency in English and a computer device with a functioning camera and microphone.

%% file: sections/04_results.tex
\section{Results}
\label{sec:results}

We collected data from 30 interaction sessions. Due to connection problems or data collection issues, we only used data from 27 interaction sessions (54 participants, a total of 3663 transcribed utterances). Participants' age ranged from $20-87$ years ($M\pm SD: 42.33\pm15.78$). Out of the 54 participants, 28 identified as female, 25 as male, one as non-binary. Race/ethnicity was mostly Caucasian/White (41), followed by Asian/Asian American (5), Hispanic/Latino (5), African/African American/Black (3), Middle Eastern/North African (1) and American Indian (1) (participants could select multiple). Most participants were native speakers of English (49), with 5 proficient users.

\subsection{Personality and demographics}

To investigate \textit{RQ1}, we evaluated whether personality or demographic traits impact how users are perceived with an ANCOVA. We examined the effects of each \textbf{personality trait} (agreeableness, conscientiousness, extroversion, neuroticism, openness) on both \textbf{Interpersonal Perception (IP)} and \textbf{Interpersonal Agreeableness (IA)}, while controlling for \textbf{gender} and \textbf{age}. The effect of \textit{agreeableness} on \textbf{IP} (the slope of the cumulative ratings curve, as evaluated by each discussion partner) was significant ($F(1,53)=6.67,p=0.01$). 
\textit{Agreeableness} was also a predictor of \textbf{IA} (how the participant came across during an interaction, reported via survey by their discussion partner), $F(1,53)=4.48,p=0.04$, as well as \textit{neuroticism} ($F(1,53)=4.65,p=0.04$).

\subsection{Conversational balance}

\begin{figure}[h] 
    \centering
    \includegraphics[width=0.43\textwidth]{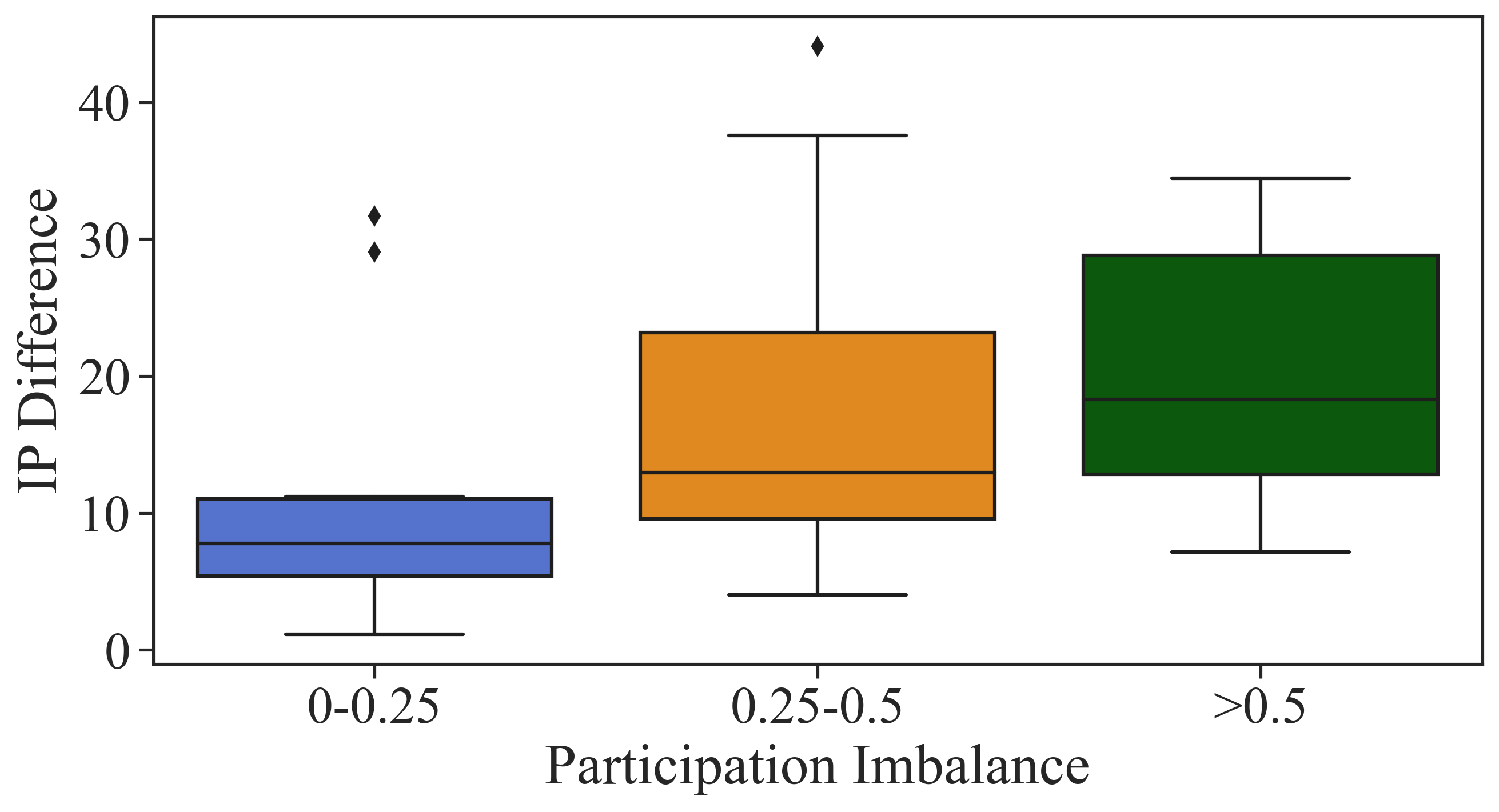}
    \caption{Conversational Imbalance impacts the alignment between the Interpersonal Perceptions of the two participants. More imbalanced conversations correlate with dissimilar retrospective ratings (higher IP Difference). Imbalance was discretized for the purpose of visualization ($N= 11$ for imbalance $<0.25$, $N= 10$ for imbalance $<0.5$, $N= 6$ for imbalance $>0.5$).}
    \label{fig:ip}
\end{figure}

We considered how aspects of conversational balance can impact differences in the perception of the interaction. For this, we considered whether the difference in the \textbf{IP} values for the dyad ($|IP_1 - IP_2|$), as well as the \textbf{IA value difference}, were impacted by the \textit{imbalance} in the conversation. We also evaluated the effects of the difference in the \textit{total number} and \textit{average length} of turns for both participants. An ANCOVA revealed that conversational \textbf{imbalance} is a predictor of the \textbf{IP difference} ($F(1,26)=4.79, p=0.04$). The interaction between these two variables can be seen in Fig. \ref{fig:ip}.

\subsection{Speech emotional content}

Finally, to answer \textit{RQ3}, we evaluated whether the content of the speech from each participant could impact how users are perceived. We ran an ANCOVA to evaluate the effects of the number of \textbf{agreement words}, as well as the \textbf{positive emotional tone ratio}, on \textbf{IP} and \textbf{IA}. Our results show that the ratio of positive tone utterances (in relation to all utterances from that participant) has a significant effect on the \textbf{IP} value ($F(1,54)=8.55, p=0.005$).

%% file: sections/05_discussion.tex
\section{Discussion}
\label{sec:discussion}

Our study contributes to the broader understanding of affective computing by emphasizing the significance of the perception of the other in shaping affective experiences. It expands the existing knowledge by highlighting the need to consider social distance as a key factor in human-robot interactions \cite{jung2017affective, caffi1994toward, andersen1996principles}. By incorporating such considerations into the design and development of affective computing systems, we can create more empathetic and effective interactions between humans and robots.

We investigated if the personality traits of users impact how they are perceived in an interaction, to answer \textit{RQ1}. Indeed, we found that the personality trait of \textit{agreeableness} has an effect on how the participant is rated by their discussion partner, both on the ``static'' measure of the interpersonal dimension (\textit{Interpersonal Agreement}), and on the continuous retrospective rating of the interaction (\textit{Interpersonal Perception}). This indicates that the brief discussion task was sufficient to grant an alignment between the perception of the self and of the other, in spite of some literature reporting that the two may conflict \cite{herringer1991perception}. 

In addition to the personality of each participant, we investigated if conversational dynamics impacted how users rated each other. To answer \textit{RQ2}, we found that the \textit{imbalance} in the speech time of each participant has an effect on how aligned their perceptions of each other are. We found that more imbalanced conversations lead to higher differences in the \textit{IP} values of the participants. Further analyses are needed to understand how exactly the interpersonal perceptions get distorted by the amount of participation in the discussion.

Finally, we answered \textit{RQ3} by investigating the emotional content of the speech of each participant. Interestingly, our results indicate that the ratio of utterances with positive emotional content impacts how the participant is rated by the other user. Agreeableness is associated with the expression of more positive emotions \cite{whittaker2021agreeable}, which may explain why participants came across as generally more agreeable.

In addition to improving our knowledge of interpersonal affect, our study and follow-up work can also contribute to bettering human-robot interactions -- agreeableness of users has been found to play a role in rapport building with robots \cite{Jeong2023agreeable}. By computing affect as a highly dynamic and context-dependent phenomenon, we can develop strategies to improve the perceived social distance between humans and robots. Understanding how different aspects of the interaction influence affective perceptions allows for the design of more seamless and responsive systems, fostering a sense of connection and engagement.